\documentclass[referee]{aa} % for a referee version
\usepackage{graphicx}
\usepackage{txfonts}
\begin{document}
   \title{The size and structure of the spheroid of IC 1613\thanks
{Based on observations obtained with MegaPrime/MegaCam, a joint project of CFHT and CEA/DAPNIA, at the Canada-France-Hawaii Telescope (CFHT) which is operated by the National Research Council (NRC) of Canada, the Institute National des Sciences de l'Univers of the Centre National de la Recherche Scientifique of France, and the University of Hawaii.}}

%

%   \subtitle{}
   \author{P. Battinelli
          \inst{1}
\and
           S. Demers \inst{2}
\and
        \'E. Artigau \inst{2}
         }

 %  \offprints{P. Battinelli}

   \institute{
INAF, Osservatorio Astronomico di Roma
              Viale del Parco Mellini 84, I-00136 Roma, Italia\\
              \email {battinel@oarhp1.rm.astro.it }
         \and D\'epartement de Physique, Universit\'e de Montr\'eal,
                C.P.6128, Succursale Centre-Ville, Montr\'eal,
                                Qc, H3C 3J7, Canada\\
                \email {demers@astro.umontreal.ca }
                \email{artigau@astro.umontreal.ca}
}

   \date{Received; accepted}

% \abstract{}{}{}{}{}
% 5 {} token are mandatory

   \abstract
{Nearby galaxies, spirals as well as irregulars,
 have been found to be much larger than previously believed. 
The structure of the huge spheroid surrounding dwarf galaxies could give
clues to their past gravitational history. Thanks to wide field imagers, 
nearby galaxies with diameter of dozens of arcmin can be effectively
surveyed.}
{We obtain, from the CFHT archives, a series of $i'$ and $g'$ MegaCam images 
of IC 1613 in order to determine the stellar surface density of the field
and determine the shape of its spheroid.}
{From the colour magnitude diagram we select 
some 36,000 stars, in the first three magnitudes of the red giant branch.
The spatial distribution of these stars is used to establish the structure
of the spheroid.}
{The position angle of the major axis of the stellar spheroid is found to be 
$\approx 90^\circ$, some
30$^\circ$ from the major axis of the HI cloud surrounding IC 1613. 
The surface density profile of the spheroid is not exponential over all
the length of the major axis. A King profile,
with a core radius of 4.5$'$ and a tidal radius of 24$'$ fits the data.
The tidal truncation
of the spheroid suggests that IC 1613 is indeed a satellite of M31.}
{}

\keywords{ galaxies, individual: IC 1613 -- galaxies: structure
}

\titlerunning{IC 1613}

\maketitle
%
%________________________________________________________________

\section{Introduction}
In recent years surveys of outer regions of nearby galaxies have
revealed that galaxies are surprisingly large.
Suffice to mention of few examples: the disk of NGC 300 has been 
detected to up to 10 scale lengths (Bland-Hawthorn et al. 2005);
stars belonging to M31 are found up to a distance of 165 kpc 
(Kalirai et al. 2006);
while for dwarf irregular galaxies, Battinelli et al. (2006) mapped the
spheroid of NGC 6822 to 36$'$ making it as big as the SMC. The less 
luminous dwarf, Leo A was also found to possess a huge halo/spheroid,
(Vansevi\v cius et al. 2004). For even less massive systems, 
McConnachie \& Irwin (2006b) have  
demonstrated that the dwarf spheroidals, satellites of M31, are
generally larger than previously recognized.

The recent advance in wide field imaging allows the survey of 
nearby galaxies with angular diameters reaching nearly one degree.
Such survey provides information on the individual halo
structure and permits the determination of universal parameters 
related to the morphology. With this in mind, we have selected one of the
bright dwarf irregular galaxies of the Local Group. 

Luminosity wise, with M$_v$ = --14.9, IC 1613 ranks 6$^{th}$ among the 15 
dwarf irregular (dIrr)
galaxies of the Local Group, being between NGC 3109 and
Sextans A. Its distance has been accurately determined by Dolphin et al.
(2001), from the apparent magnitude of the red giant branch tip (TRGB) and
the red clump stars, who
obtained $\mu_o$ = 24.31 $\pm$ 0.06 or 730 $\pm$ 20 kpc.
Pietrzy\'nski et al. (2006) determined its distance to be $\mu_o$ = 24.291
$\pm$ 0.014, from 
near infrared photometry of its Cepheids. Based on these two estimates, we
adopt 722 $\pm$ 5 kpc for the distance of IC 1613. 
 Its global structure has been investigated by Hodge et al.
(1991) using UK Schmidt plates scanned with COSMOS. They follow the 
radial stellar density profile to 400$''$. They were, however, unable to
decide if an exponential profile, with a scale length of 800 pc or a
King model with a core radius of 200 $\pm$ 50$''$ and a tidal radius
of 1420 $\pm$ 25$''$ give a better representation of the profile.
The major axis diameter of IC 1613 is given by NED/IPAC to be 16.2$'$.
This diameter appears small on the light of the 
 carbon star survey by Albert et al. (2000) who identified some 200
C stars extending to 15$'$, essentially doubling the size of IC 1613.
Tikhonov \& Galozutdinova (2002) later mapped the surface density of
giants ($8'< r<12'$) and concluded that the density variation follows and
exponential and that they have not reached the limit.

The HI density and velocity maps of IC 1613 were analyzed by Lake \&
Skillman (1989). They conclude that the rotation curve of IC 1613 shows
evidences for dark matter in the outer parts. They established that the
position angle of the major axis is 58 $\pm 4^\circ$. 
A survey to fainter limits by
Hoffman et al. (1996) reach a 
maximum diameter of $54'$. 
Since Hodge et al. (1991) optical survey barely  reached 7$'$ from the center, 
it is certainly warranted to survey a larger area to establish which 
surface density profile best fits the data.

\section{Photometric data}

We use for our survey five $i'$ and five $g'$ archival 
CFHT Megacam images.
The wide field imager MegaCam consists of 36 2048 $\times$ 4612 pixel
CCDs, covering nearly a full 1$^\circ \times 1^\circ$ field. It is mounted
at the prime focus of the 3.66~m Canada-France-Hawaii Telescope. 
It offers a  resolution of 0.187 arc second per pixel.
These filters are designed to match
Sloan Digital Sky Survey (SDSS) filters, as defined by the Smith et al.
(2002) standards.  The exposure times were, respectively 205 s and
135 s. Ten consecutive exposures were taken, essentially all with
the same airmass. 

The images secured from the CFHT archives were already processed and calibrated by 
Elixir (Magnier \& Cuillandre 2004). The Elixir system performs a pre-reduction
of the Megacam mosaic to produce calibrated flat-fielded, fringe-corrected images
ready for further processing. In order to obtain a uniform zero point across the
mosaic,  all 36 CCDs are converted to a similar gain times quantum efficiency.
This is easily achieved by dividing the mosaic image by a flat-fielded Mosaic
which CCDs have been normalized with respect to CCD00 to maintain the ratio of
gain and quantum efficiency. The result of the division is a flat looking image.
\footnote{For further details see:
http://www.cfht.hawaii.edu/Instruments/Imaging/MegaPrime/dataprocessing.html}
Magnier \& Cuillandre (2004) quote a formal scatter of magnitude residuals, 
across the mosaic of 0.0086 mag. 

These images, obtained in Queue observing mode, are provided with calibration
equations. However, since the calibration includes colour terms which are
function of ($g'-r'$) or ($r'-i'$) we cannot fully calibrate our data
because 
 there is no $r'$ observations avavilable for IC 1613.
No aperture correction has been applied to the derived photometry.
 Taking into account the
exposure times and the airmasses, the calibration equations given
in the headers can be expressed as:

$$ g' = g'_{inst} + 6.781 + 0.15(g'-r'),\eqno(1)$$
$$ i' = i'_{inst} + 6.520 + 0.08(r'-i'),\eqno(2)$$

where the instrumental magnitudes refer to those calculated by DAOPHOT.
The exposure times are sufficiently short that we have acquired only the 
first few magnitudes of the stellar population of IC 1613, as we shall 
display in the next Section. Since most
stars are on the red giant branch (RGB), with ($g'-r') \approx$ 1.3,
while 0.6 $<$ $(r'-i')$ $<$ 1.6, 
the $g'$ colour term may be as large as 0.2 mag, or 
even larger for carbon stars which have extreme $(g'-r')$ colours;
 while the $i'$ colour term may reach 0.13 mag. The exact magnitude
and colour are not essential for our goals. For the moment we simply
neglect the colour terms and take into account only the magnitude
zero points.

\subsection{Data analysis}

Before precessing the images with DAOPHOT we astrometrically combine 
the five exposures of each filter into one $g'$ and $i'$ image for each of the 
36 CCDs of the MegaCam mosaic. 
The DAOPHOT-II/ALLSTAR package 
(Stetson 1987, 1994) is selected to process each CCD image. Thresholds are 
adopted, employing Stetson's procedure. Final instrumental magnitudes
are obtained with  ALLSTAR which fits a model PSF to all stellar objects on the 
frame. Figure 1 displays the photometric errors, as defined by ALLSTAR,
 as a function of the $i'$
magnitudes for one CCD which includes IC 1613. 
%____________________________________________Fig. 1 sigma distribution
   \begin{figure*}
 \centering
\includegraphics[width=8cm]{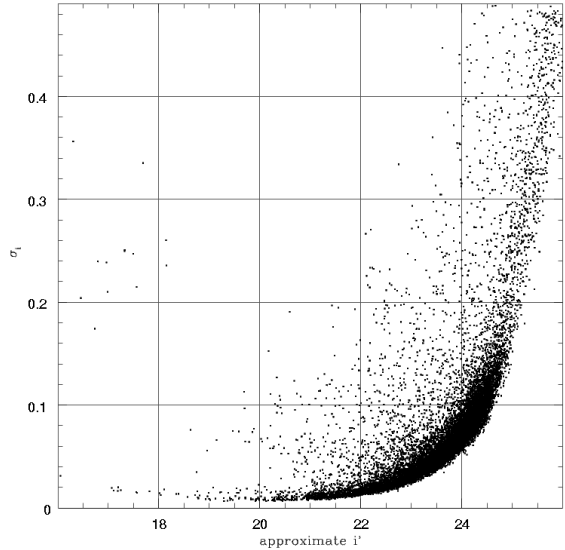}

 \caption{The photometric error, as determined by ALLSTAR, for a CCD which
includes IC 1613. To reduce the colour scatter we exclude objects with
magnitude error, both in $i'$ and $g'$, less than 0.15.
}
             \label{Fig sigma}
   \end{figure*}
%---------------------------------------------------------

In order to exclude objects with large colour errors we select, 
before combining the
$i'$ and $g'$ magnitude files, only stars with photometric 
errors (as defined by ALLSTAR)
 less than 0.15 mag.
This results in a total of 85809 objects in one square degree.
Because
IC 1613 is located at a Galactic latitude of b = $-61^\circ$, an impressive
number of galaxies is seen in the surrounding field. Therefore, to reduce
the background counts these non-stellar objects should be eliminated.
DAOPHOT-II provides an image quality diagnostics SHARP that can be used
to separate stellar and non-stellar objects. For isolated
stars, SHARP should have a value close to zero, whereas for semi-resolved 
galaxies
and unrecognized blended doubles SHARP will be significantly greater than zero.
On the other end, bad pixels and cosmic rays produce SHARP less than zero.
SHARP
must be interpreted as a function of the apparent magnitude of all objects
because the SHARP parameter distribution flares up near the magnitude limit;
see Stetson \& Harris (1988) for a discussion of this parameter. 

Figure 2 shows the SHARP distributions, as a function of the apparent 
magnitude, for two CCDs inside and two CCDs outside of IC 1613.
The dashed line, at 16.80 mag, indicates the saturation limit.
We see a large population of non-stellar
objects well above the limiting magnitude of the sample. 
Indeed, visual inspection of several $i'$ and $g'$ images reveals that 
bright objects with larger SHARP have larger than normal FWHM while fainter ones
are obviously non-stellar. 
Based on this
figure, we exclude objects with $\vert$SHARP$\vert$ $>$ 0.5 from 
our stellar database.
There are 63000 stars satisfying the magnitude and SHARP criteria, their spatial
distribution, in the one squared degree area, is presented in Figure 3. The 
gaps between the Megacam CCDs are evident. Bright foreground stars are responsible for 
the small regions lacking stars. Just North-East of
IC 1613 the empty spot is due to an 8th magnitude star.

%____________________________________________Fig. 2 sharp distributions
   \begin{figure*}
 \centering
\includegraphics[width=9cm]{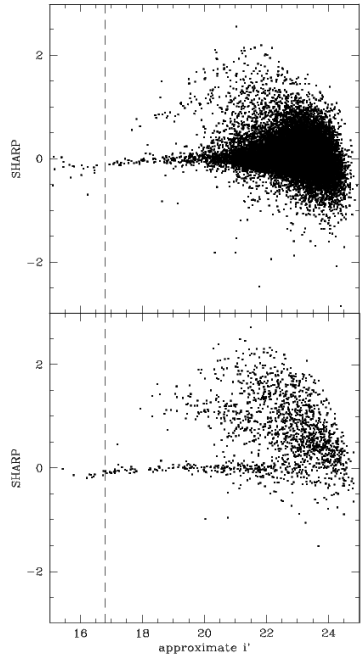}

 \caption{Distributions of the SHARP parameter for objects with
photometric error $<$ 0.15 mag. The upper panel corresponds to two CCDs within
IC 1613 while the bottom panel is for two CCDs outside the galaxy. 
We see a huge population of non-stellar
objects brighter that the magnitude limit. Close to the magnitude limit,
SHARP becomes meaningless due to the large scatter. The dashed line marks
the saturation threshold.
}
             \label{Fig sharp}
   \end{figure*}
%---------------------------------------------------------

\subsection{Completeness tests}

In order to establish the efficiency of DAOPHOT/ALLSTAR to recover faint stars among
the various stellar density encountered, we perform a series of completeness
tests. For a given CCD of the mosaic, the test consists in the addition of 140 stars, uniformly distributed
in magnitude between $i'$ = 20.0 and 24.6 and with corresponding $g'$ magnitudes
that match giant branch stars of IC 1613. 
The stars are added to the $i'$ and $g'$ images,
they are then analyzed with DAOPHOT/ALLSTAR, the results are combined to produce
a list of stars with magnitudes and colours, thus only stars with recovered 
$i'$ and $g'$ are counted. 
As representative of different crowding conditions  over the field, we selected 
 2 in the center, 2 in the periphery and 2 outside the galaxy. 	For each CCD we
performed 6 tests with a total of 980 added stars.
We find that, except for the two central CCDs where
the density is highest, the completeness is 75\% 
at $i'$ = 24.0 and drops to 50\% 
and 24.6. The completeness of the two central CCDs  at $i'$ = 24.0 is 50\%. 
However, because of the strong radial gradient of the surface density, 
the completeness in the 
inner halves of the central CCDs is even lower while in the outer halves we find
completeness levels ($\sim $80\%)
 similar to those found for the periphery. For the 
above reasons, we will exclude the inner 5$'$ from any further interpretation 
of the data.    
In the following analysis we restrict our counts to stars
brighter than $i'$ = 24.0.  

%____________________________________________Fig. 3  plot of all stars
   \begin{figure*}
 \centering
\includegraphics[width=8cm]{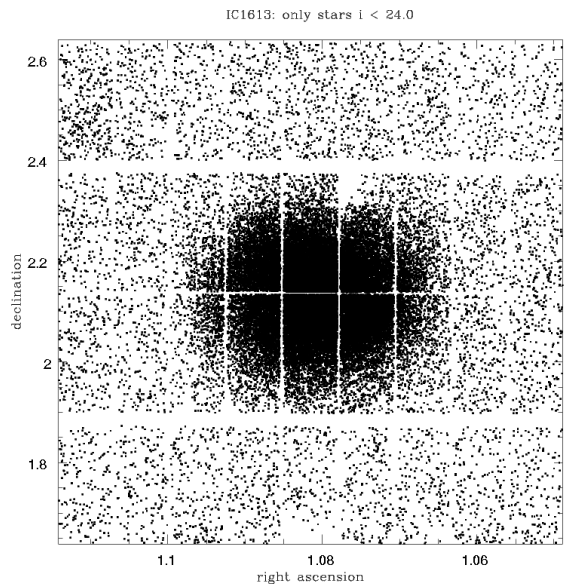}

 \caption{Spatial distribution of all stars with $i'$ $<$ 24.0 and selected
following the SHARP criterion described above.
}
             \label{Fig RA plot}
   \end{figure*}
%---------------------------------------------------------

\section{Results}
\subsection{Space distribution of young and old populations}

Figure 4 presents the colour-magnitude diagram of the whole $1^\circ
\times 1^\circ$ field. Only stars with error and SHARP satisfying the
adopted criteria are plotted. A well-defined giant branch is seen,
along with main sequence stars and an extended AGB where known C stars
are located. According to Dolphin et al. (2001), the 
TRGB of IC 1613 is seen at I = 20.35 $\pm 0.07$. On Fig. 4
we see that the number of stars per magnitude interval nearly double
between  $i'$ = 20.80 and 20.90. This is essentially consistent with
the quoted TRGB since the difference between the $i'$ and I magnitudes
in the colour range of the TRGB is $\sim$0.60 mag (Battinelli et al. 2006).

%____________________________________________Fig. 4 combined CMD
   \begin{figure*}
 \centering
\includegraphics[width=8cm]{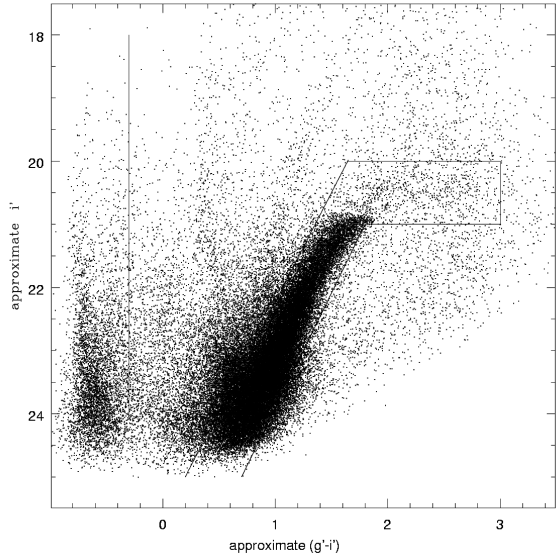}

   \caption{Colour-magnitude diagram of the whole field, 63,000 objects
satisfying the error and SHARP criteria, are plotted. The giant box encloses
the RGB and AGB stars, 
while the blue stars are in the left region. The white line
marks the 75\%
 completeness level for the RGB stars that will be used for 
the structural study of the galaxy.
}
              \label{FigCMD}
    \end{figure*}
%---------------------------------------------------------

%____________________________________________Fig. 5 non sharp CMD
   \begin{figure*}
 \centering
\includegraphics[width=8cm]{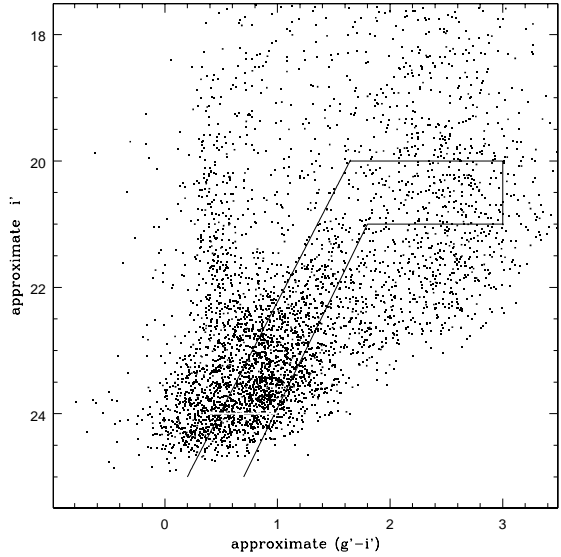}

   \caption{Same as Fig. 4 but for objects rejected because of their
high SHARP.}
              \label{FigCMD}
    \end{figure*}
%---------------------------------------------------------
As can be seen in the CMD of Figure 5, the presumably non-stellar objects, 
rejected because of their high SHARP, show a wide range of colours.
One thousand non-stellar objects, with low reddening selected from the
SLOAN database, albeit brighter than those seen here, show a huge 
colour spread. They have a mean $(g' - i')$ of 1.8 $\pm 1.4$

To investigate the distribution of individual stars, in the plane of the sky,
we transform their equatorial coordinates into Cartesian x,y coordinates.
We adopt, for the origin: x = 0, y = 0, 
 the coordinates of IC 1613 as given by NED/IPAC, namely:
$\alpha$ = 01:04:48.8 and  $\delta$ = 02:07:04.0 (J2000.0).
Stars fainter than $i'$ = 20.0 can be divided into two groups: the blue
main sequence stars having $(g'-i') < -0.3$ and the RGB + AGB stars, in the
box drawn on Fig. 4.
The spatial distributions of these  two populations  differ, as can be
seen in Figure 6. There are 5000 main sequence stars and 36,500 RGB +
AGB stars in this figure. Giants define an elliptical spheroid whose
major axis is nearly oriented East - West. The younger main sequence stars 
are concentrated in an area close to the center but show an off center
asymmetry to the center defined by the giants. They also appear to have
an elliptical distribution  oriented unlike the spheroid. The outer 
contour and three representative contours of the HI map of IC 1613 
obtained by Hoffman et al. (1996) are
traced. Lake \& Skillman (1989) give 58 $\pm 4^\circ$ for the position
angle of the major axis of the HI envelope. 
The RGB stars are nearly all withing the outer isodensity contour.
In Fig. 6 (top panel) we also show for comparison the spatial
distribution of blue stars. We are awarre that, even though dominated by 
upper-MS stars, this sample certainly includes older blue supergiants.
The spatial distribution of blue stars shows a clear offset relative to the
HI and RGB stars. 
 
%_____________________________________________Fig.6  spatial distributions
   \begin{figure*}
   \centering
\includegraphics[width=9cm]{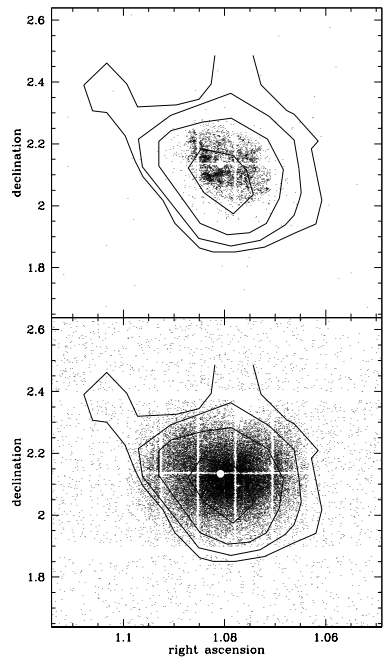}

   \caption{Spatial distributions of the blue main sequence stars (top
panel) and RGB + AGB stars (lower panel). 
Representative HI isodensity contours, from Hoffman et al. (1996) are
traced.  Each panel corresponds to one square degree of the sky. Big white 
dots indicate the position of the calculated center.
}
              \label{maps}
    \end{figure*}

\subsection{Star counts and shape of IC 1613}
The gaps between the MegaCam CCDs are far from negligible
and need to be filled to better define the stellar surface density 
distributions. 
 The two wide horizontal empty strips, seen in Fig. 3 above and below the
galaxy, 
($\approx 80''$ wide) are the largest gaps in the mosaic.
The narrower vertical gaps (13$''$) are more numerous but not as critical. Lacks of stars here and there are the 
results of very bright stars contaminating their surrounding. 
The horizontal gaps are filled by duplicating the observed stellar
distribution on a strip of width 40$''$ above and 40$''$ below the gap.
Similar fill-up is done for the vertical gaps.
Such gap-filling was previously used on MegaCam data by Battinelli et al. (2006).
The whole field is then covered by a 50 $\times$ 50 pixel wide grid
and stars are counted  over a circular 500 pixel sampling area, centered
on each intersection of the grid. This is done to  smooth out major
irregularities (mainly due to bright foreground stars that locally prevent
the detection of fainter members of IC 1613.

%_____________________________________________Fig.7 PA's & Elliptiicties
   \begin{figure*}
   \centering
\includegraphics[width=8cm]{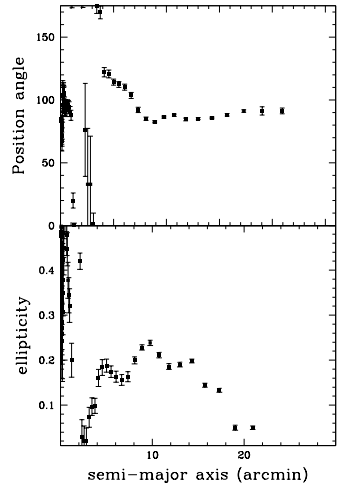}

   \caption{Solutions for the position angles and ellipticities for 
elliptical isodensity contours.
}
              \label{parameters}
    \end{figure*}
%---------------------------------------------------------
The density map is then
transformed into a density image that is analyzed with IRAF/STSDAS/ELLIPSE
task to fit isodensity ellipses, determine their position angles and
ellipticities. The results of this task are summarized in Figure 7 where
we show the position angles (PA) and the ellipticities ($\epsilon$) derived
for semi-major axes. The PA and $\epsilon$, as expected,
 vary wildly close to the centre where the stellar surface density is quite
patchy and irregular. For distances larger than 4$'$ their 
variations become much less erratics. 

For the purpose of the following analysis, we approximate the PA and $\epsilon$
as following: PA = $135^\circ$ for r $< 4'$, decreases linearly to 
PA = $87^\circ$ at r = $7'$, then stays constant for r $> 7'$. Since $\epsilon$ is
ill defined for r $< 4'$, we assume that is its value is equal to its mean
value between $4'$ and $12'$, namely 0.19 $\pm$ 0.02. 
for r $< 12'$, it decreases linearly toward zero when reaching r = $22'$.
The surface density of the elliptical annuli, as defined above, is displayed in
Figure 8. 
The observed density reaches
a plateau at r $\sim 23'$, we adopt for the foreground + background contributions
to the giant counts a density of 0.659 $\pm$ 0.073 ``giants'' arcmin$^{-2}$.
This is determined from the mean of the last six points.
This number is subtracted from the observed counts, to obtain the corrected
surface density.

%_____________________________________________Fig. 8 profile
   \begin{figure*}
   \centering
\includegraphics[width=8cm]{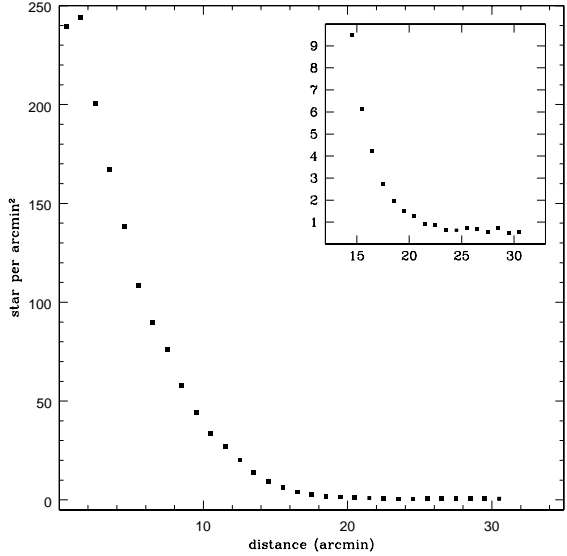}

   \caption{The observed density profile of the giants. A close up of the tail
is shown in the small panel.
}
              \label{profile}
    \end{figure*}
%---------------------------------------------------------

The centers of the outer ellipses ($ 10' < r < 20'$), as determined by 
the IRAF task,
 does not match exactly to the origin of the coordinates. Indeed, the mean of
the centers  is located at a position $48 \pm 8''$ East
and $58 \pm 14''$ North of the adopted NED/IPAC origin. Therefore, the center
of IC 1613 spheroid is then: $\alpha$ = $1^h$ 4$^m$ 51.0$^s$, $\delta$ =
$2^\circ$ 8$'$ 02$''$ (J2000).

\subsection{Density profile of the spheroid}

Figure 9 displays ln$\rho$, (the corrected surface density) versus the 
semi-major axis of the ellipses. The error bars are based on the number of
stars counted in each annulus.
The figure reveals that the density profile deviates from a straight line,
excluding, of course, the inner $\sim 5'$ where we know that the incompleteness
of our counts is more severe.
Contrary to the conclusion of Tikhonov \& Galazytdinova (2002) who mention
that IC 1613 shows an exponential density profile up to 12$'$, we observe a 
pronounced decline to well beyond $20'$. 
This implies that a simple exponential profile does not represent well
the whole observed profile. 
A more complex function, involving more free parameters,
must then be sought.
King models (King 1962) have been fitted to globular clusters and dSph 
galaxies (Irwin \& Hatzidimitriou 1995; Caldwell et al 1992; McConnachie
\& Irwin 2006b). 
For more massive galaxies the 
S\'ersic (1968) law has been considered, see for example Binggeli \& Jerjen
(1998). We shall compare the observed profile to both King and S\'ersic models.
 
%_____________________________________________Fig.6 ln profile
   \begin{figure*}
   \centering
\includegraphics[width=7cm]{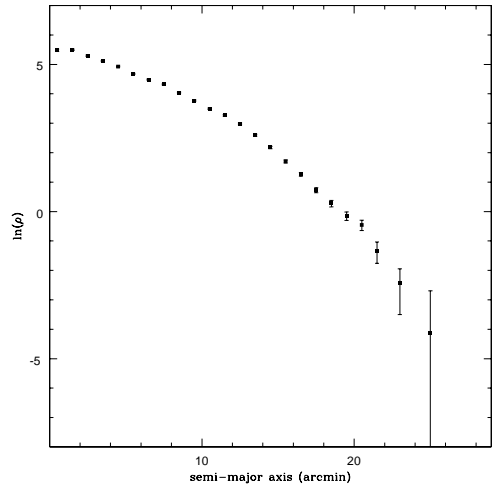}

   \caption{The corrected surface density profile of the spheroid, defined by
its giants, is not exponential.
}
              \label{ln profile}
    \end{figure*}

\subsubsection{The King profile}

The fit  of King's equation to the corrected density profile for $r > 5'$ 
is displayed in Figure 10. 
The adopted fit yields: 
a corrected central density k = 596 $\pm$ 168 star 
arcmin$^{-1}$, a core
radius of $r_c$ = 4.54 $\pm 0.93'$ and a tidal radius of $r_t$ = 24.4 
$\pm 0.3'$. For the adopted distance of IC 1613, $r_c$ = 953 $\pm$ 195 pc
and $r_t$ = 5.12 $\pm$ 0.06 kpc. This tidal radius is nearly identical to the
one deduced by Hodge et al. (1991).
The spheroid of 
IC 1613 is obviously bigger than all known dSphs but its
concentration parameter, defined as: c = log$_{10}({r_t\over r_c})$, is
c = 0.7 a typical value found among dSphs. 
As expected, our best fitted model gives an overestimate for the central region 
(dashed line in  Fig. 10).
Indeed, as discussed in Sect. 2.2, the completeness factor in 
the central region is significantly lower 
than elsewhere in the field.

%_____________________________________________Fig.10 King profile
   \begin{figure*}
   \centering
\includegraphics[width=7cm]{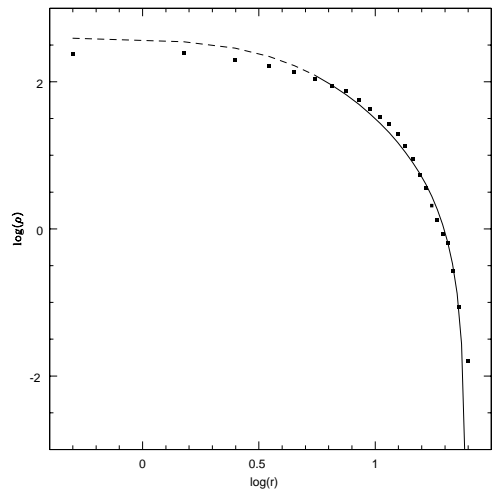}

   \caption{A King model is fitted to the surface density of giants in the 
spheroid. Dashed line is an extrapolation of the fit. Error bars are omitted to better
show the dispersion of the points along the curve.
}
              \label{King profile}
    \end{figure*}
%---------------------------------------------------------

\subsubsection{The S\'ersic profile}

S\'ersic law, more general than an exponential profile, is normally 
applied to surface brightness profiles. It can also be used for the
surface stellar density profile. It is expressed in the 
following way:

$$ \rho = \rho_oe^{{-(r/b)}^n},\eqno(3)$$
It includes a third parameter, $n$, which controls the shape of the profile.
Note that many authors use $1/n$ rather than $n$. 
A fit to our density profile for $r>5'$ yields: $\rho_o$ = 179.6 $\pm$ 7.9, $b =
7.94 \pm 0.21'$ (1.64 kpc) and $n$ = 1.83 $\pm$ 0.07. The curve corresponding to these
parameters is displayed in Figure 11. The value of $n$ is rather large and
unusual, indeed, none of the galaxies investigated by Binggeli \& Jerjen
(1998) have such a large value. The parameter $n$ was found by Caon et al. (1993)
to be function of the effective radius ($b$). With a $b = 1.64$ kpc, IC 1613 falls
off the observed trend because its $n$ is too large. Finally, we note also 
that the S\'ersic fit underestimates the observed central density, this is 
obviously unrealistic.  

%_____________________________________________Fig. 11 Sersic profile
   \begin{figure*}
   \centering
\includegraphics[width=7cm]{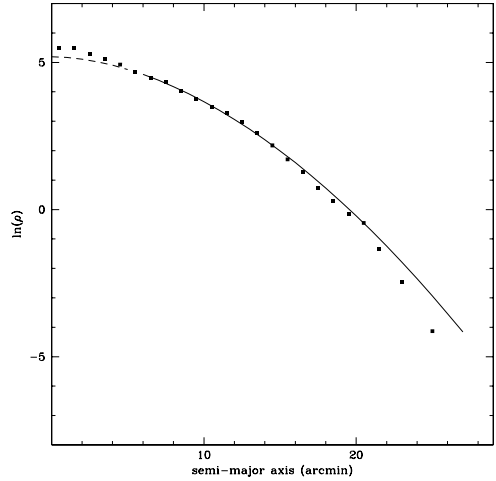}

   \caption{A S\'ersic law fitted to the observed profile for $r>5'$. Dashed line
is an extrapolation of the fit.
}
              \label{Sersic profile}
    \end{figure*}
%---------------------------------------------------------

  \section{Discussion}

At 510 kpc from M31, IC 1613 is currently closer to Andromeda 
than to the Milky Way. The physical
association of IC 1613 with M31 is, however, a matter of debate.
Mateo (1998) explains that IC 1613 is an ambiguous case that could be
associated with M31 or the MW; while  van den Bergh (2000) considers IC 1613 to
be a free floating object, isolated in the Local Group. 
In their recent investigation of the satellite distribution of M31,
McConnachie \& Irwin (2006a) conclude that the large distance of IC 1613 to
M31 makes its membership to the M31 subgroup marginal, see their Fig. 2.
Adopting a different point of view, 
for their mass estimate of M31, Evans \& Wilkinson (2000) include 
IC 1613 among its satellites, on the ground that it is nearer to M31 than to the
Milky Way. This question will not be fully solved until the space velocity 
of IC 1613 relative to M31 is determined. 

We believe that the King model gives an appropriate representation of the 
profile of the spheroid of IC 1613 contrary to the S\'ersic profile,
 traced in Fig. 11. 
The observed sharp decline of the profile of the spheroid 
militates in favor of a gravitational association with its
nearest big neighbour. However, as often stated, (e.g. McConnachie \& Irwin 2006b)
a King model is not physically motivated for dSphs or more massive galaxies
since their relaxation time is of the order of the Hubble time and their
stellar velocity distribution may deviate significantly from Maxwellian. 
Even if Fornax, the most massive dSph associated with
the Milky Way, has a density profile following amazing well a King model
(Walcher et al. 2003) 
 any interpretation of the structure of such galaxies based
exclusively on the King model fit must be treated with caution. 

Notwithstanding this caveat, we could  then evaluate, 
from the determined tidal radius,
the distance of nearest approach of IC 1613 to M31.  Satellite galaxies are
expected to be tidally truncated to the theoretical tidal radius at 
perigalacticon (Oh \& Lin 1992). 
To do so, we first have to adopt a mass for each galaxy. Contrary
to dSphs devoid of gas, IC 1613 has a dynamical mass determined from its
HI, we take  
$8\times 10^8$ M$_\odot$ (Mateo 1998) and $2\times 10^{12}$ 
M$_\odot$ for M31 (Evans \& Wilkinson 2000). 
The simple point mass approximation must
certainly be skipped for M31. Indeed, its halo is huge having a radius 
of at least 165 kpc (Kalirai et al. 2006). One other alternative
is to adopt a logarithmic potential following 
Oh et al. (1995) formulation linking the tidal radius (r$_t$) of the satellite, 
the eccentricity ($e$) end the semi-major axis (a) of its orbit and 
the mass ratio ($\alpha$), equal to $4\times 10^{-4}$ in our case.
The three variables are linked in the following way:
$$ r_t = \left[{\alpha(1 - e^2)}\over{[(1-e)^2/2e]{\rm ln}[(1+e)/(1-e)] +1}\right]^{1/3}{\rm a} \eqno(4)$$

The above equation rules out a circular orbit for IC 1613. 
When $e\rightarrow 0$
eq. 4 becomes $r_t = {\rm a}\alpha^{1/3}$ leading to a semi-major axis of
69 kpc, substantially less than the current distance of IC 1613 to M31.
However, the application of equation 4 with  eccentricities from 0.0 to 0.95
yields a solution for $e = 0.78$ and $a = 290$ kpc.
Such ellipse  has a apogalacticon at 517 kpc, close to the actual distance
of IC 1613 to M31. The perigalacticon would then be 63 kpc, which is
certainly a reasonable minimum distance. However, such large orbit  
around M31 has a period of  $\sim$10 Gyr, that is nearly one Hubble time! 
Such hypothetical orbit is obviously meaningless because the metric of the
Local Group has increased appreciably over the last 10 Gyr. 

\section{Conclusion}
Our wide field survey of IC 1613 
 confirms Tikhonov \& Galazutdinova (2002) finding that no giants of
IC 1613 are seen in their HST observations at $27'$ and $33'$ from its center.
The stellar component of IC 1613 is found to extend as far as its observed
HI envelope. We note that the young stellar population has a different
spatial distribution than the old stars defining the spheroid. 
The orientation of the HI cloud is appreciably different from
the orientation of the stellar spheroid. More pronounced differences between
the morphology and orientation of the stellar and gaseous components are
seen in NGC 6822, another Local Group dIrr
(Battinelli et al. 2006). Nevertheless, IC 1613 dual morphology raises
intriguing questions concerning the origin if the HI cloud.

Contrary to the isolated NGC 6822, the spheroid of IC 1613 shows a mark
truncation which matches  a King density profile. We attribute this feature
to the fact that IC 1613 belongs to the M31 system. 

\begin{acknowledgements}
This research
is funded in parts (S. D.) by the Natural Sciences and Engineering Research
Council of Canada. 
\end{acknowledgements}

\end{document}